\documentclass[pdflatex,sn-mathphys-ed,iicol]{sn-jnl}

\usepackage{graphicx}
\usepackage{dcolumn}
\usepackage{bm}
\usepackage{hyperref}
\usepackage{lipsum}
\usepackage{slantsc}
\usepackage{slashed}
\usepackage{mathtools}
\usepackage{siunitx}
\usepackage{xcolor}
\usepackage[T1]{fontenc}
\usepackage{lmodern} 
\usepackage{upgreek}
\usepackage{amsfonts}
\usepackage{amsmath}
\usepackage{amssymb}
\rmfamily 

\usepackage{mathtools, cuted}
\usepackage{balance}

\usepackage{xcolor}
\definecolor{color3}{RGB}{51,51,153}

\DeclareFontShape{T1}{lmr}{b}{sc}{<->ssub*cmr/bx/sc}{}
\DeclareFontShape{T1}{lmr}{bx}{sc}{<->ssub*cmr/bx/sc}{}

\DeclarePairedDelimiterXPP\BigOSI[2]%
  {\mathcal{O}}{(}{)}{}%
  {\SI{#1}{#2}}

  \usepackage{xcolor}
\definecolor{color3}{RGB}{51,51,200}

\hypersetup{
     colorlinks=true,
     linkcolor=color3,
     urlcolor=color3,
     }

\jyear{2025}

\theoremstyle{thmstyleone}

\theoremstyle{thmstyletwo}

\theoremstyle{thmstylethree}

\begin{document}

\title[]{Minicharged Particle Sensitivity of the MAPP Outrigger Detector}

\author[1]{\fnm{Matti} \sur{Kalliokoski}}

\author[2]{\fnm{Vasiliki A.} \sur{Mitsou}}

\author[3]{\fnm{Marc} \sur{de Montigny}}

\author[4]{\fnm{Abhinab} \sur{Mukhopadhyay}}

\author[5]{\fnm{Pierre-Philippe A.} \sur{Ouimet}}

\author[4]{\fnm{James} \sur{Pinfold}}

\author[4]{\fnm{Ameir} \sur{Shaa}}

\author*[2,4]{\fnm{Michael} \sur{Staelens}}\email{\href{mailto:michael.anthony.staelens@cern.ch}{michael.anthony.staelens@cern.ch}}

\affil[1]{\orgdiv{Helsinki Institute of Physics}, \orgname{University of Helsinki}, \orgaddress{\street{Gustaf H\"allstr\"omin katu 2}, \postcode{00014} \city{Helsinki}, \country{Finland}}}

\affil[2]{\orgdiv{Instituto de F\'isica Corpuscular}, \orgname{CSIC--Universitat de Val\`encia}, \orgaddress{\street{Catedr\'atico Jos\'e Beltr\'an, 2}, \postcode{46980} \city{Paterna}, \country{Spain}}}

\affil[3]{\orgdiv{Facult\'e Saint-Jean}, \orgname{University of Alberta}, \orgaddress{\street{8406 Rue Marie-Anne Gaboury}, \city{Edmonton}, \state{AB} \postcode{T6C 4G9}, \country{Canada}}}

\affil[4]{\orgdiv{Department of Physics}, \orgname{University of Alberta}, \orgaddress{\street{11335 Saskatchewan Drive NW}, \city{Edmonton}, \state{AB} \postcode{T6G 2E1}, \country{Canada}}}

\affil[5]{\orgdiv{Department of Physics}, \orgname{University of Regina}, \orgaddress{\street{3737 Wascana Parkway}, \city{Regina}, \state{SK} \postcode{S4S 0A2}, \country{Canada}}}

\abstract{We present a detailed study of the projected background-free sensitivity of the MAPP Outrigger Detector (OD) to minicharged particles (mCPs) at the High-Luminosity Large Hadron Collider (HL-LHC). As the first upgrade to the MAPP Experiment, the MAPP OD is a standalone detector designed to offer enhanced sensitivity to high-mass mCPs with intermediate effective charges. The MAPP OD is planned for installation in a duct adjacent to the MAPP-1 detector, located between the LHC's UA83 gallery and the beamline. Considering mCP production via the Drell--Yan mechanism and various meson decays, the results show that, at the 95\% confidence level, the MAPP OD can extend the experiment's upper mass reach to mCP masses of approximately 200~GeV at the HL-LHC.}


\maketitle

\section{Introduction}
\label{Sec:Intro}
The Standard Model (SM) describes nature's basic elements; it comprises all the constituents of normal matter and details their properties and interactions. Although it has successfully predicted  numerous phenomena and explained most experimental results with remarkable precision, there is a growing consensus that it is incomplete. For example, dark matter, strongly supported by multiple independent lines of evidence~\cite{1933AcHPh, 1937ApJ86, 1970ApJ159, COBE1, COBE2, Komatsu_2009, Komatsu_2011, Planck2015cosmo}, remains unexplained by the SM. Dark matter is simply another form of matter; thus, a similar particle-based description as normal matter seems natural, yet no particle candidate has been found. This presents one of today's greatest unsolved puzzles in physics.

Weakly interacting massive particles (WIMPs) have arguably been the leading candidate; however, over $40$~years of rigorous searches have yielded no discovery. Consequently, comprehensive searches for other well-motivated candidates are necessary. A compelling alternative is feebly interacting particles (FIPs)~\cite{Lanfranchi2021, Antel2023}, which emerge naturally in dark sector models. These models feature a new hidden or dark sector that minimally couples to the SM visible sector via portal interactions~\cite{Patt2006,Batell2009,Beacham2020,Lanfranchi2021}. Within this framework, minicharged particles (mCPs) can arise through kinetic mixing between a dark $U(1)$ gauge field and the SM hypercharge gauge field~\cite{Holdom1986_1,Holdom1986_2}. These mCPs acquire small effective electric charges and thus interact feebly with ordinary matter, offering a unique window into dark sector physics.

Minicharged particles have been invoked in addressing anomalies such as the EDGES 21~cm hydrogen absorption signal~\cite{Bowman2018,Munoz2018,Berlin2018}. Over the past four decades, extensive experimental efforts have placed stringent constraints on the mCP parameter space, yet substantial regions remain unexplored, particularly at the energy frontier accessible at the Large Hadron Collider (LHC).

The MoEDAL Apparatus for Penetrating Particles (MAPP) is a dedicated LHC experiment designed to probe this unexplored territory~\cite{StaelensPhDThesis,pinfold2023moedalmapp}. In particular, mCPs constitute the flagship benchmark scenario for the Phase-1 MAPP detector, MAPP-1~\cite{Kalliokoski2024}. This paper provides a follow-up to our previous work~\cite{Kalliokoski2024}, presenting the first projected sensitivity results of the MAPP Outrigger Detector (OD)---the experiment's first upgrade---to mCPs at the HL-LHC.

The paper is organized as follows. In Section~\ref{Sec:Outrigger}, we describe the design and construction of the MAPP OD in detail. Section~\ref{Sec:mCP_Prod} covers the modeling and simulation of mCP production at the LHC, considering a variety of production mechanisms. In Section~\ref{Sec:Results}, we present the projected sensitivity of the MAPP OD to mCPs, including an associated scenario involving minicharged strongly interacting dark matter (mC-SIDM). Finally, conclusions and future prospects are provided in Section~\ref{Sec:Conc}.

\section{The MAPP Outrigger Detector}
\label{Sec:Outrigger}
\begin{figure}[htb]
	\centering 
	\includegraphics[width = 7.5 cm]{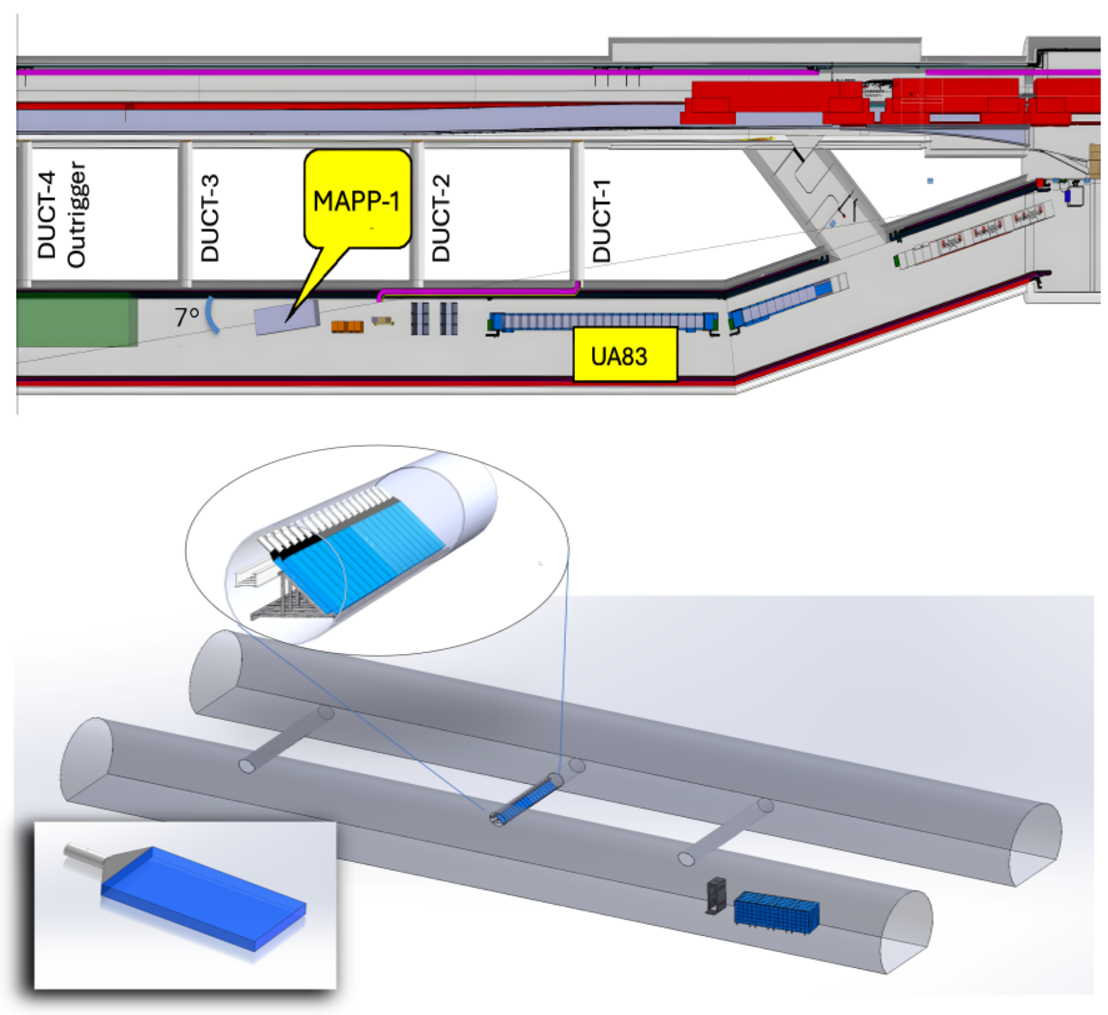}			
        \caption{\textbf{Upper:} An overview of the UA83 gallery, indicating the location of Duct~4 relative to IP8, the beamline, and the MAPP-1 detector. \textbf{Lower:} A schematic diagram illustrating the design and placement of the MAPP Outrigger Detector in Duct~4. The bottom-left panel shows a basic scintillator plank used in the MAPP Outrigger Detector; MAPP-1 is displayed in the bottom-right of the diagram.} 
	\label{fig:fig_ORschematics}
\end{figure}

The MAPP OD is an auxiliary detector for the MAPP-1 experiment, specifically aimed at improving the experiment's sensitivity to high-mass, intermediate-charge mCPs. The MAPP OD is proposed for installation in Duct~4 of the UA83 tunnel, adjacent to the MAPP-1 detector, approximately $120$~m from Interaction Point~8 (IP8). The ducts are numbered sequentially, starting with the duct closest to IP8. Figure~\ref{fig:fig_ORschematics} provides an overview of the UA83 region, highlighting the placement of the ducts in relation to the MAPP-1 detector. Duct~4 has been approved for use by MoEDAL-MAPP, while Ducts~1--3 are reserved for the machine group, and the status of Duct~5 remains uncertain. The choice of duct has minimal impact on the MAPP OD's sensitivity~\cite{Pinfold:2918254}. 

The MAPP OD comprises $80$ plastic scintillator slabs of size $60$~cm $\times$ $30$~cm $\times$ $5$~cm (Bicron BC-408) arranged in a layered ``bricklayer'' configuration. This design allows for coincidence-based event selection, improving background rejection and signal identification, and offers self-veto capabilities. Each scintillator slab is coupled to a single low-noise photomultiplier tube (PMT); the first three detector layers use $2^{\prime\prime}$ Hamamatsu R2154-02s, while the fourth layer uses $3.5^{\prime\prime}$ HZC Photonics XP82B2FNBs.

The design concept of the MAPP OD follows the principle of a hodoscope array. The basic installation subunit comprises eight scintillator slabs arranged in four layers on a rail. The units will be mounted onto the rail at a $45^{\circ}$ angle to increase the scintillator path length for through-going particles originating from the IP, extending it from $5$~cm to $\sim7$~cm. Grouping eight scintillator slabs into a single installation subunit is designed to ease manual handling and enable event selection through signal coincidences across the layers. In total, ten of these basic installation subunits will be deployed within Duct~4 to construct the MAPP OD. An additional layer of iron shielding will be positioned between the MAPP OD and the duct opening facing the beamline.

The MAPP OD PMT readout and calibration system will be integrated into the MAPP-1 detector's main electronics rack. Both the readout and calibration systems will mirror the functional design of MAPP-1, described in more detail in Refs.~\cite{Kalliokoski2024,Pinfold:2918254,Pinfold2791293}.

The MAPP OD addition to the MoEDAL-MAPP facility is currently under review by the LHC Experiments Committee.

\section{Minicharged Particle Production at the LHC}
\label{Sec:mCP_Prod}
In this work, we follow our modeling and analysis approach described in detail in Ref.~\cite{Kalliokoski2024}, which includes mCP production via a variety of mechanisms: 1) the Drell--Yan mechanism; 2) direct decays of vector mesons; and 3) single Dalitz decays of pseudoscalar mesons. Although non-exhaustive, these production modes lead to abundant mCP production at the LHC. This section reviews the key components and details of our methodology.

\subsection{\label{subsec:DYprod}Drell--Yan Pair Production}
The Drell--Yan production of minicharged particles (mCPs) was simulated using a dedicated \textsc{FeynRules}~\cite{CHRISTENSEN2009,Christensen2011} model implementation. The corresponding \textsl{Universal \textsc{FeynRules} Output} (UFO)~\cite{DEGRANDE2012} model was processed by \textsc{MadGraph5}\_aMC@NLO (MG5; v2.7.3)~\cite{Alwall2011,Alwall2014} to perform the simulations. The leading-order pair production of mCPs via the Drell--Yan mechanism was simulated at the parton level. Drell--Yan production with inclusive hadronization will be addressed in a future, complete Monte Carlo study that also incorporates detailed mCP energy loss spectra, detector response, and background modeling. Nevertheless, we predict favorable kinematics relative to the placement of the MAPP-1 detector and MAPP OD; thus, the sensitivity projections associated with Drell--Yan production reported in this study are expected to be conservative by comparison. 

Following Holdom's model that originally predicted mCPs~\cite{Holdom1986_1}, wherein an additional massless $U(1)$ gauge field $A'_{\mu}$ kinetically mixes with the SM hypercharge gauge field $B^{\mu}$, our implementation features the following Lagrangian:
\begin{align}
    \mathcal{L} &= \mathcal{L}_{\mathrm{SM}} -\frac{1}{4}  A'_{\mu \nu} A'^{\mu \nu} \nonumber \\
    &\quad + i \bar{\chi} \left( \slashed{\partial} + i e' \slashed{A}' - i \kappa e' \slashed{B} + i m_{\chi} \right) \chi, \label{EQN:Lag}
\end{align}
where $A'_{\mu \nu}$ is the dark photon field strength tensor following the usual definition of $A'_{\mu \nu} = \partial_{\mu} A'_{\nu} - \partial_{\nu} A'_{\mu} $; $\chi$ is a new Dirac fermion with mass $m_{\chi}$ coupled to the dark photon gauge field $A'_{\mu}$, and hence charged under it with an electric charge, $e'$; and $\kappa$ is an arbitrary (potentially irrational) small parameter that governs the kinetic mixing strength.

The expected number of mCPs $\left( N_{\chi} \right)$ produced at IP8 through Drell–Yan pair production during a given LHC run can be estimated as follows:
\begin{equation}
    N_{\chi} \simeq 2 \sigma_{q\bar{q} \rightarrow \chi \bar{\chi}} L^{\mathrm{int}}_{\mathrm{LHCb}},  \label{EQN:NmCP_DY}
\end{equation}
where $\sigma_{q\bar{q} \rightarrow \chi \bar{\chi}}$ represents the Drell--Yan pair-production cross-section, and $L^{\mathrm{int}}_{\mathrm{LHCb}}$ is the estimated integrated luminosity at IP8 for a given LHC run, assumed hereafter to be $300$~fb$^{-1}$ for the HL-LHC.

\subsection{\label{subsec:Mdec}Meson Decays}
mCPs can be copiously produced through the decays of various mesons, proceeding directly via their electromagnetic coupling to the photon. In analogy with Eq.~\ref{EQN:NmCP_DY}, the total yield of mCPs from meson decay processes can be estimated as follows:
\begin{equation}
    N_{\chi} \simeq 2   \mathcal{B}_{M\rightarrow \chi \bar{\chi} X } \sigma_{M}  L^{\mathrm{int}}_{\mathrm{LHCb}},   \label{EQN:NmCP_MDec}
\end{equation}
where $\mathcal{B}_{M\rightarrow \chi \bar{\chi} X }$ denotes the branching ratio of the meson decay channel producing mCP pairs and $\sigma_{M}$ is the meson's production cross section. The branching ratios for both direct and Dalitz decays contributing to mCP production are presented in detail in the following subsections.

\subsubsection{\label{subsec:Dir}Direct Decays of Vector Mesons}
Electromagnetic decays of neutral vector mesons (such as $\rho$, $\omega$, $\phi$, $J/\psi$, $\psi\left(2S\right)$, $\Upsilon\left(nS\right)$) can directly produce mCP pairs, with the corresponding branching ratio $\mathcal{B}_{M\rightarrow \chi \bar{\chi}}$ derived by scaling the meson's electron--positron decay branching ratio ($\mathcal{B}_{M \rightarrow e^{-} e^{+}}$) by $\epsilon^{2}$ and an mCP-mass-dependent phase-space correction~\cite{Kelly2019}. Accordingly, this can be expressed as
\begin{equation}
    \mathcal{B}_{M\rightarrow \chi \bar{\chi}} = \epsilon^{2} \mathcal{B}_{M \rightarrow e^{-} e^{+}} I^{(2)} \left(\frac{m_{\chi}^2}{m_{M}^{2}},\frac{m_{e}^2}{m_{M}^{2}}\right),   \label{EQN:BRdir1}
\end{equation}
where $m_{e}$ and $m_{M}$ represent the electron and parent meson masses, respectively, and $I^{(2)}\left(x,y\right)$ is the following function describing the two-body decay~\cite{Kelly2019}:
\begin{equation}
    I^{(2)}\left(x,y\right) = \frac{\left(1+2x\right)\sqrt{1-4x}}{\left(1+2y\right)\sqrt{1-4y}}.     \label{EQN:I2}
\end{equation}
This yields the following expression for the branching ratio of neutral vector mesons decaying directly into mCP pairs:
\begin{equation}
    \mathcal{B}_{M\rightarrow \chi \bar{\chi}} = \epsilon^{2} \mathcal{B}_{M \rightarrow e^{-} e^{+}} \frac{\left( m_{M}^{2} + 2m_{\chi}^{2} \right) \sqrt{m_{M}^{2}-4m_{\chi}^2}}{\left(m_{M}^{2} + 2m_{e}^{2} \right) \sqrt{m_{M}^{2} - 4m_{e}^{2}}}.   \label{EQN:BRdir2}
\end{equation}

\subsubsection{\label{subsec:Dal}Dalitz Decays of Pseudoscalar Mesons}
Single Dalitz decays of neutral pseudoscalar mesons (such as $\pi^0$, $\eta$, $\eta'$) can produce a pair of mCPs and a photon. We do not consider double Dalitz decays, as they are strongly suppressed by $\epsilon^{4}$. The branching ratio for neutral pseudoscalar mesons undergoing single Dalitz decays into mCPs is given by
\begin{equation}
    \mathcal{B}_{M\rightarrow \gamma \chi \bar{\chi}} = \epsilon^{2} \alpha \mathcal{B}_{M \rightarrow \gamma \gamma} I^{(3)}\left(\frac{m_{\chi}^{2}}{m_{M}^{2}} \right),
\end{equation}
where $\mathcal{B}_{M \rightarrow \gamma \gamma}$ represents the corresponding SM branching ratio to two photons, and $I^{(3)}(x)$ is the following function describing the three-body decay~\cite{Kelly2019}:
\begin{equation}
    I^{(3)}(x) = \frac{2}{3\pi} \int_{4x}^{1} dz \sqrt{1 - \frac{4x}{z}} \frac{\left(1-z\right)^{3}\left(2x+z\right)}{z^2}.
\end{equation}
This yields the following expression for the branching ratio of neutral pseudoscalar mesons undergoing single Dalitz decays to mCP pairs:
\begin{strip}
\begin{equation}
\label{EQN:BRDal}
    \mathcal{B}_{M\rightarrow \gamma \chi \bar{\chi}} = \epsilon^{2} \alpha \mathcal{B}_{M \rightarrow \gamma \gamma} \left(\frac{2}{3\pi} \int_{4m_{\chi}^{2}/m_{M}^{2}}^{1} dz \sqrt{1 - \frac{4m_{\chi}^{2}/m_{M}^{2}}{z}} \frac{\left(1-z\right)^{3}\left(2m_{\chi}^{2}/m_{M}^{2}+z\right)}{z^2}\right).
\end{equation}
\end{strip}

All meson masses and SM branching ratios used in our calculations were obtained from the 2022 PDG review~\cite{PDG2022}. To evaluate Eq.~\ref{EQN:BRDal}, numerical integrations were performed in \textit{Mathematica} (v11.2); the value of the fine-structure constant was set as $\alpha = 1/137$. The leading-order Drell--Yan production cross-section was estimated directly using our MG5 model with the default NNPDF2.3QED parton distribution function~\cite{Ball2013} and $\alpha\left(m_{Z}\right) \simeq 1/127.94$~\cite{PDG2022}. For heavy quarkonia production, $\sigma_{M}$ was estimated using cross-sections and multiplicities from $pp$ collisions simulated using \textsc{Pythia}~8 (v8.240)~\cite{SJOSTRAND2015159,Bierlich2022} with the default Monash 2013 tune~\cite{Skands2014}. For light vector and pseudoscalar mesons, cross-sections were normalized to the inelastic $pp$ cross-section, taken as $79.95$~mb for the HL-LHC ($\sqrt{s}=14$~TeV)~\cite{Aaboud2016}. All equations and calculations were validated through comparisons with similar results reported in the literature, as detailed in Ref.~\cite{Kalliokoski2024}.

\section{Sensitivity of the MAPP Outrigger Detector to Minicharged Particles}
\label{Sec:Results}
In line with our previous study~\cite{Kalliokoski2024}, the projected sensitivity of the MAPP OD to mCPs was established by estimating the expected signal event yield ($N_{\mathrm{sig}}$) as follows:
\begin{equation}\label{Eqn:Nsig}
    N_{\mathrm{sig}} = N_{\chi} \times A \times P
\end{equation}
where $A$ is the detector's geometric acceptance to mCPs produced by a given process; $N_{\chi}$ is the number of mCPs produced by such a process, estimated via Eq.~\ref{EQN:NmCP_DY} or \ref{EQN:NmCP_MDec}; and $P$ is the detection probability based on the expected photoelectron yield from a through-going mCP. Monte Carlo simulations were performed to estimate the detector acceptance as a function of the mCP mass for each production mode described in Sec.~\ref{Sec:mCP_Prod}. Specifically, an mCP was only considered ``accepted'' if it traversed all four collinear sections of the MAPP OD. For heavy quarkonia production, we followed the suppression scheme outlined in Ref.~\cite{Foroughi2021}, implementing the following suppression factor in \textsc{Pythia}~8 via the \texttt{SuppressSmallPT} user hook:
\begin{equation}\label{Eqn:pTsupp}
    \frac{p_{\mathrm{T}}^{4}}{\left(\left(k p_{\mathrm{T}0} \right)^{2} + p_{\mathrm{T}}^{2} \right)^{2}} \left( \frac{ \alpha_{\mathrm{s}}\left( \left(  k p_{\mathrm{T}0} \right)^2 + Q^2_{\mathrm{ren}} \right)}{\alpha_{\mathrm{s}}\left( Q^2_{\mathrm{ren}} \right)} \right)^{n} , \nonumber
\end{equation}
where $Q_{\mathrm{ren}}$ is the renormalization scale, and $p_{\mathrm{T}0}$ and $p_{\mathrm{T}}$ are the energy-dependent dampening scale and transverse momentum, respectively. Following Ref.~\cite{Foroughi2021}, values of $k = 0.35$ and $n = 3$ were selected.

The detection probability for a through-going mCP in an $n$-layer scintillation detector is given by the following Poisson distribution:
\begin{equation}
    P=\left(1 - {\rm e}^{-N_{\mathrm{PE}}}  \right)^{n}.
\end{equation}
Here, $N_{\mathrm{PE}}$ is the number of detected photoelectrons; for mCPs, $N_{\mathrm{PE}}$ scales approximately as $\epsilon^2 N_{\gamma} \mathrm{QE}$, with $N_{\gamma}$ representing the number of scintillation photons reaching the PMT, and $\mathrm{QE}$ representing the PMT's mean quantum efficiency, taken to be $25\%$. \textsc{Geant4}~\cite{AGOSTINELLI2003} (v10.6.p02) photon yield simulations were conducted to estimate $N_{\gamma}$ for a minimally ionizing particle traversing a MAPP OD scintillator plank unit. The following parameters were used in our simulations: 1) a surface reflectivity of $98\%$; 2) a light output of $10000$~$\gamma$/MeV; 3) a bulk light attenuation length of $2.6$~m; 4) scintillator plank dimensions of $60$~cm $\times$ $30$~cm $\times$ $5$~cm; and 5) a CSA light guide with dimensions $5$~cm $\times$ $7.5$~cm and an index of refraction of $1.44$. Simple Tyvek\textsuperscript{\textregistered} wrapping from default \textsc{Geant4} libraries was also included. Photon transport was handled using the UNIFIED~\cite{Nayar1991,Levin1996} and LUT Davis models~\cite{Roncali2013,Roncali2017}. For $1$~GeV muons, simulations indicate an average of $N_{\gamma} \simeq 2.488 \times 10^4$ optical photons reaching the PMT.

\begin{figure}[!htb]
	\centering 
 	\includegraphics[width = 7.5 cm]{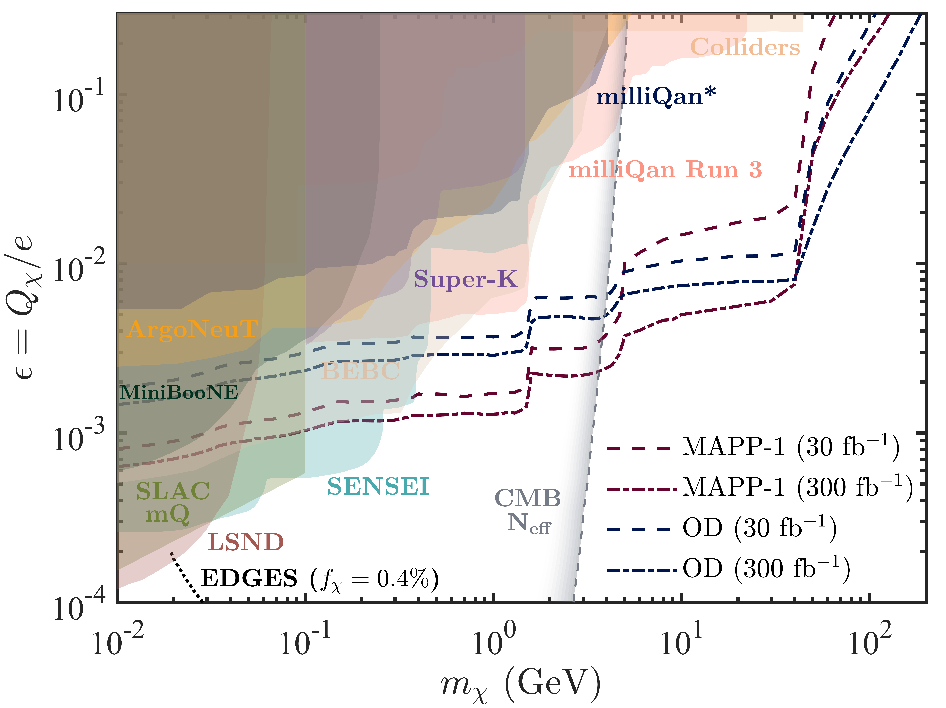}	
        \caption{The projected $95\%$ CL exclusion limits for the MAPP Outrigger Detector (blue) for minicharged particles vs. those for the MAPP-1 detector (red)~\cite{Kalliokoski2024}. The dashed lines correspond to $L^{\mathrm{int}}_{\mathrm{LHCb}} = 30$~fb$^{-1}$ and the dashed-dotted lines correspond to $L^{\mathrm{int}}_{\mathrm{LHCb}} = 300$~fb$^{-1}$. The shaded areas represent the regions of parameter space excluded by previous searches at the $95\%$ CL (except the BEBC bounds, which correspond to the $90\%$ CL)~\cite{Prinz1998,Davidson2000,Magill2019,Acciarri2020,Ball2020,Plestid2020,Marocco2021,Barak2024,alcott2025search}; milliQan* denotes the milliQan demonstrator~\cite{Ball2020}. The grey dashed and black dotted lines represent the most stringent indirect $2\sigma$ upper limits derived from the Planck full-mission results~\cite{Planck2018cosmo} on the effective number of different neutrino species ($N_{\mathrm{eff}}$)~\cite{Adshead_2022} and the region associated with a potential resolution of the EDGES anomaly ($f_{\chi} = 0.4\%$)~\cite{Kovetz2018}, respectively.} 
	\label{fig:fig_OR_mCP_lims}
\end{figure}

Using the $CL_{s}$ method~\cite{Read2002} and assuming a background-free scenario, the $95$\% confidence level (CL) exclusion limits are defined as regions where at least three signal events are expected~\cite{PDG2022}. The corresponding projected $95\%$ CL sensitivity of the MAPP OD for integrated luminosities of $30$ and $300$~fb$^{-1}$, is presented in Figure~\ref{fig:fig_OR_mCP_lims}. The $300$~fb$^{-1}$ curve corresponds to the projected integrated luminosity at IP8 for the HL-LHC, while the $30$~fb$^{-1}$ curve is included for comparison with one-tenth the luminosity. The figure also shows a comparison between the projected sensitivity of the MAPP OD and that of the MAPP-1 detector previously reported in~\cite{Kalliokoski2024}. In both panels, previous experimental constraints on the mass--mixing parameter space~\cite{Prinz1998,Davidson2000,Magill2019,Acciarri2020,Ball2020,Plestid2020,Marocco2021,Barak2024,alcott2025search} are shown, along with indirect $2\sigma$ upper limits derived from CMB measurements of the effective number of neutrino species ($N_{\mathrm{eff}}$)~\cite{Adshead_2022}. The region of parameter space relevant to an mCP-based explanation of the EDGES anomaly, assuming a maximal minicharged dark matter fraction of $f_{\chi}=0.4\%$ allowed by Planck~2015 CMB constraints~\cite{Planck2015cosmo,Planck2015cosmoII} is also included~\cite{Kovetz2018}. The results demonstrate that, at the $95\%$ confidence level, the MAPP OD can extend the MAPP Experiment's sensitivity to mCP masses up to approximately $200$~GeV at the HL-LHC.

\subsection{Minicharged Strongly Interacting Dark Matter}
Minicharged strongly interacting dark matter (mC-SIDM)~\cite{Mahdawi2018,Emken2019,Plestid2020,Foroughi2021,Alexander2021,KLING2023} presents an alternative dark matter scenario in which a small subcomponent of dark matter interacts more strongly with SM particles than is typically considered in WIMP searches. Unlike traditional dark matter candidates that are assumed to interact weakly, mC-SIDM can possess interaction cross-sections sufficiently large to leave observable signals in detection media through processes such as elastic dark matter--electron scattering. However, the sensitivity of terrestrial direct-detection experiments is fundamentally limited; above a critical cross-section ($\bar{\sigma}_{e\mathrm{,ref,crit}}$), mC-SIDM particles undergo sufficiently frequent interactions in the atmosphere and Earth's crust that they lose substantial energy before reaching underground detectors, rendering their signals undetectable~\cite{Emken2018}. This critical cross-section, which defines an upper bound of experimental sensitivity for direct detection experiments, cannot be overcome by ground-based searches and motivates the exploration of alternative detection strategies, such as high-altitude and accelerator-based experiments. 

Although dark matter composed entirely of minicharged particles is strongly excluded, cosmological data still allow for a small minicharged subcomponent of up to $f_{\chi} \leq 0.4$\% of the total dark matter abundance. Within this framework, several experiments, including terrestrial direct-detection experiments~\cite{Mahdawi2018,Emken2019}, a rocket-borne search~\cite{Erickcek2007}, and a high-altitude balloon-based experiment~\cite{Rich1987}, have set stringent limits on mC-SIDM. Nonetheless, a distinct region of the mC-SIDM parameter space remains unconstrained and currently accessible to accelerator experiments, presenting unique opportunities for future exploration. Notably, accelerator-based searches provide a significant advantage in probing this scenario, as their sensitivity does not depend on the assumed fraction of minicharged dark matter.

In the mC-SIDM scenario, exclusion plots are typically presented using the following reference cross-section for DM--electron scattering:
\begin{equation}\label{Eqn:RefXS}
    \bar{\sigma}_{e\mathrm{,ref}} = 16 \pi \alpha^{2} \epsilon^{2} \mu^{2}_{\chi e}/q^{4}_{d\mathrm{,ref}},
\end{equation}
where $\mu_{\chi e}$ denotes the $\chi$--electron reduced mass, and $q_{d\mathrm{,ref}}$ represents the reference three-momentum transfer, set as $q_{d\mathrm{,ref}} = \alpha m_{e}$ following Refs.~\cite{Essig2012,Emken2019,Foroughi2021}.

\begin{figure}[htb]
	\centering 
	\includegraphics[width = 7.5 cm]{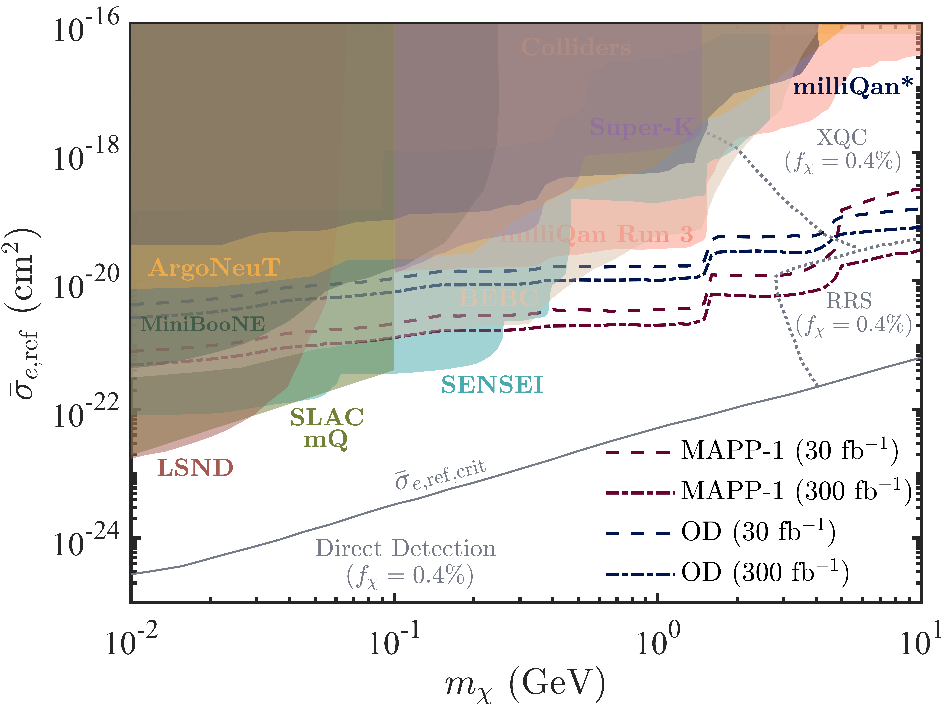}
    \caption{The projected $95\%$ CL exclusion limits for the MAPP Outrigger Detector (blue) for mC-SIDM vs. those for the MAPP-1 detector (red)~\cite{Kalliokoski2024}. The dashed lines correspond to $L^{\mathrm{int}}_{\mathrm{LHCb}} = 30$~fb$^{-1}$ and the dashed-dotted lines correspond to $L^{\mathrm{int}}_{\mathrm{LHCb}} = 300$~fb$^{-1}$. Additional exclusion limits shown on the plot in grey correspond to projections of constraints set by several strongly interacting dark matter searches assuming a small minicharged fraction of dark matter of $f_{\chi} = 0.4$\%: terrestrial direct-detection experiments~\cite{Mahdawi2018,Emken2019}, the X-ray quantum calorimetry (XQC) experiment~\cite{Erickcek2007}, and a high-altitude balloon-based experiment (RRS)~\cite{Rich1987}. The solid grey line denotes the critical reference cross-section ($\bar{\sigma}_{e\mathrm{,ref,crit}}$); milliQan* denotes the milliQan demonstrator~\cite{Ball2020}.} 
    \label{fig:fig_OR_mCSIDM_lims}
\end{figure}

By applying Eq.~\ref{Eqn:RefXS} to reinterpret existing constraints and our projected exclusion limits on mCPs from the MAPP OD in the context of the mC-SIDM scenario, we obtain the results shown in Figure~\ref{fig:fig_OR_mCSIDM_lims}. A comparison of these projections with those for the MAPP-1 detector reported in Ref.~\cite{Kalliokoski2024} is provided for integrated luminosities of $30$ and $300$~fb$^{-1}$. The figure also includes the additional exclusion bounds associated with SIDM searches assuming a minicharged dark matter subcomponent fraction of $f_{\chi} = 0.4$\%~\cite{Mahdawi2018,Emken2019,Erickcek2007,Rich1987}. Both detectors demonstrate sensitivity to previously unconstrained regions of the mC-SIDM parameter space.

\section{Conclusions}
\label{Sec:Conc}
This study presented the latest results from our investigation of the MAPP OD in the mCP benchmark scenario. We considered multiple mCP production mechanisms in $pp$ collisions, including the Drell--Yan process, direct decays of vector mesons, and single Dalitz decays of pseudoscalar mesons. The results indicate that the MAPP OD can improve the sensitivity of the MAPP-1 experiment to higher mass mCPs with intermediate effective charges, extending the upper mass reach of the experiment to approximately $200$~GeV. Moreover, in the context of the minicharged strongly interacting dark matter scenario, the MAPP OD offers complementary sensitivity to a distinct and unexplored region of parameter space that remains inaccessible to conventional direct-detection experiments. These results build on the foundation established by MAPP-1, indicating the MAPP OD's potential in advancing the search for mCPs at the LHC within the MAPP Experiment's physics program.

\balance

\backmatter

\bmhead{Acknowledgments}
We are grateful to the Natural Sciences and Engineering Research Council of Canada (NSERC) for partial financial support: Discovery Grant, SAPPJ-2019-00040; Research Tools and Instruments Grants, SAPEQ-2020-00001 and SAPEQ-2022-00005. M.d.M. thanks NSERC for partial financial support under Discovery Grant No. RGPIN-2016-04309. M.S. acknowledges support by the Generalitat Valenciana (GV) via the APOSTD Grant No. CIAPOS/2021/88. V.A.M. and M.S. acknowledge support by the GV via Excellence Grant No. CIPROM/2021/073, as well as by the Spanish MCIN/AEI/10.13039/501100011033/ and the European Union/FEDER via the Grant PID2021-122134NB-C21. We also thank the authors and maintainers of the \textsc{FeynRules}, \textsc{MadGraph}~5, \textsc{Pythia}~8, and \textsc{Geant}4 software packages.

\section*{Statements and Declarations}

\bmhead{Data Availability Statement}
The results reported in this study were obtained from analyses of simulation data; no original datasets were created.

\bmhead{Competing Interests}
The authors declare no competing or conflicting interests.

\bmhead{Author Contributions}
Conceptualization, J.P. and M.S.; methodology, M.K., M.d.M., A.M., P.-P.A.O., and M.S.; validation, A.M., A.S., and M.S.; formal analysis, M.K. and A.M. (\textsc{Geant}4), and M.S. (\textsc{MadGraph}~5, \textsc{Pythia}~8); investigation, M.S.; resources, V.A.M. and J.P.; data curation, M.S.; writing---original draft preparation, M.S.; writing---review and editing, V.A.M., P.-P.A.O., J.P., and M.S.; visualization, J.P. and M.S.; supervision, V.A.M. and J.P.; project administration, J.P.; funding acquisition, V.A.M., M.d.M., J.P., and M.S. All authors have read and agreed to the published version of the manuscript.

\bibliography{Minicharge_A}

\end{document}